\documentclass[prl,twocolumn,showpacs,superscriptaddress,twoside,floatfix]{revtex4}

\usepackage{amssymb,amsmath}
\usepackage{graphicx}
\usepackage{color}
\usepackage{bm}

\begin{document}

\title{Enhanced Transmission of Terahertz Radiation through Periodically Modulated Slabs of Layered Superconductors}

\author{D.V.~Kadygrob}
\affiliation{A.Ya.~Usikov Institute for Radiophysics and Electronics, National Academy of Sciences of Ukraine, 61085 Kharkov, Ukraine}

\author{N.\,M.\,Makarov}
\affiliation{Benem\'erita Universidad Aut\'{o}noma de Puebla, Puebla, Pue. 72000, M\'{e}xico}

\author{F. P\'erez-Rodr\'{\i}guez}
\affiliation{Benem\'erita Universidad Aut\'{o}noma de Puebla, Puebla, Pue. 72000, M\'{e}xico}

\author{T.M. Slipchenko}
\affiliation{A.Ya.~Usikov Institute for Radiophysics and Electronics, National Academy of Sciences of Ukraine, 61085 Kharkov, Ukraine}

\author{V.A.~Yampol'skii \footnote{yam@ire.kharkov.ua}}
\affiliation{A.Ya.~Usikov Institute for Radiophysics and Electronics, National Academy of Sciences of Ukraine, 61085 Kharkov, Ukraine}
\affiliation{Benem\'erita Universidad Aut\'{o}noma de Puebla, Puebla, Pue. 72000, M\'{e}xico}
\affiliation{V.N.~Karazin Kharkov National University, 61077 Kharkov, Ukraine}

\begin{abstract}
We predict the enhanced transmissivity of modulated slabs of layered superconductors for terahertz radiation due to the diffraction of the incident wave and the resonance excitation of the eigenmodes. The electromagnetic field is transferred from the irradiated side of a slab of layered superconductor to the other one by excited \emph{waveguide modes} (WGMs) which do not decay deep into the slab, contrary to metals, where the enhanced light transmission is caused by the excitation of the \emph{evanescent surface waves}. We show that a series of resonance peaks (with $T\sim1$) can be observed in the dependence of the transmittance $T$ on the varying incidence angle $\theta$, when the dispersion curve of the diffracted wave crosses successive dispersion curves for the WGMs.
\end{abstract}

\date{\today}


\maketitle

Since the first observation by Ebbesen \emph{et al.}~\cite{ebb}, the enhanced light transmission (ELT) through metal films perforated by subwavelength holes is the focus of attention for many research groups (see, e.g., reviews~\cite{Genet_2007,Garsia-Vidal_2010} and references therein). This phenomenon is observed in films with thicknesses much larger than the skin depth, and the transmission coefficient turns out to be much larger than that predicted by Bethe's theory of electromagnetic diffraction at small apertures~\cite{bet}. The ELT is related to coupling of surface plasmons resonantly excited at both sides of the perforated film. Discussion of this and some alternative mechanisms of extraordinary transmission can be found in the review by Zayats {\it et al}~\cite{Marad}. Recent interest to the aforementioned effect is due to its possible applications for light control, photovoltaics, detection and filtering of radiation in visible and far-infrared frequency ranges.

For observation of the ELT in thick enough metal films, there should obviously exist a mechanism for the transfer of the electromagnetic energy from the irradiated side of the film to the other. In regards, we would like to mention two forgotten ways to make metal films transparent for electromagnetic waves. Both of them are characteristic for pure metals at low temperature. The first one is the so called \emph{anomalous penetration of the electromagnetic field} deep into the film by the chains of the Larmor electron orbits (see, e.g., Ref.~\cite{kaner1}). In the external dc magnetic field parallel to the sample surface, the electrons with long enough mean free path carry the electromagnetic field out from the skin layer and form an additional current layer at the distance of the Larmor diameter from the sample surface. The another group of Larmor electrons carry the electromagnetic field deeper into the sample and form the next current layer, and so on. Thus, the Larmor electrons can serve as carriers for the transfer of the electromagnetic energy through the metal films.

The second mechanism is related to different kinds of weakly decaying electromagnetic waves: helicons, dopplerons, cyclotron waves, etc. (see, e.g. Refs.~\cite{kaner2,kaner3,platzman}). Under definite special conditions, in the presence of the external dc magnetic field, they can propagate in a metal and carry the electromagnetic energy from the irradiated side of a film to the other.

In this Letter, we consider a novel mechanism for the ELT, which is a specific combination of the two mechanisms discussed above. We predict and analytically study the enhanced transmission of terahertz radiation through slabs of layered superconductors with periodically modulated values of the maximum density $J_c$ of the $\mathbf{c}$-axis Josephson current. Such a modulation can be realized either by irradiating a standard $\rm Bi_2Sr_2CaCu_2O_{8+\delta}$ sample covered by a modulated mask~\cite{kwok} or by pancake vortices controlled by an out-of-plane magnetic field~\cite{koshelevprl}. Similarly to the case of the ELT in metals, the modulation results in the generation of diffracted waves, which, under the resonance conditions, excite the electromagnetic eigenmodes.

The Josephson current along the crystallographic $\mathbf{c}$-axis couples with the electromagnetic field inside the insulating dielectric layers, forming Josephson plasma waves (JPWs) (see, e.g., review~\cite{Thz-rev} and references therein). Thus, the propagation of electromagnetic waves through the layers is favored by the layered structure. The study of these waves is very important because of their terahertz frequency range, which is still hardly reachable for electronic and optical devices. It is necessary to emphasize a principal difference of the JPWs in the strongly anisotropic Josephson plasma of layered superconductors from the electromagnetic waves in isotropic media, e.g., in standard metals. Contrary to waves in metals, the dispersion law for JPWs has a
``hyperbolic'' form~\cite{Thz-rev},
\begin{equation}\label{hyp}
\frac{k^2_{s\,x}\lambda_c^2}{\omega^2/\omega_J^2 -1}-k^2_{s\,z}\lambda_{ab}^2=1.
\end{equation}
Here $\omega$ is the wave frequency; $k_{s\,x}$ and $k_{s\,z}$ are the components of the wave vector along and across the superconducting layers; $\lambda_c =c/\omega_J\varepsilon^{1/2}$ and $\lambda_{ab}$ are the magnetic-field penetration depths along and across the layers, respectively; $\omega_J\sim1$~THz is the Josephson plasma frequency; $\varepsilon$ is the interlayer dielectric constant; $c$ is the speed of light.

Equation \eqref{hyp} shows that JPWs can propagate across the layers for $\omega>\omega_J$ only. Moreover, the longitudinal component $k_{s\,x}$ of the wave vector should be large enough,
\begin{equation}\label{kx}
k_{s\,x}>k_c=\frac{1}{\lambda_c}\sqrt{(\omega/\omega_J)^2-1}.
\end{equation}
This specific feature of the JPWs is a reason for the principal difference of ELT in metals and in layered superconductors. For metals, not only the basic wave with $k_{s\,x}=(\omega/c)\sin\theta$, but all diffracted waves with $k_{s\,x}=(\omega/c)\sin\theta+ng$, \emph{exponentially decay to the middle of a sample} (here $\theta$ is the incidence angle, $g$ is the period of the reciprocal lattice of modulations, $n$ is an integer). Contrarily, for layered superconductors, while the basic wave with $(\omega/c)\sin\theta<k_c$ exponentially decays into the sample, the diffracted waves having $|(\omega/c)\sin\theta+ng|>k_c$, \emph{can propogate across the layers} and, therefore, can serve as carriers for the electromagnetic energy, similarly to the helicons, dopplerons, and cyclotron waves described in Refs.~\cite{kaner2,kaner3,platzman}. Thus, the diffraction in metals can result in the resonant excitation of the symmetric or antisymmetric \emph{evanescent surface waves}, whereas the diffraction in the layered superconductors can provide the resonant excitation of \emph{waveguide modes} (WGMs) (see Refs.~\cite{waveguide,krokhin}), which do not decay deep into the slab but oscillate across the layers.

In this Letter, we analytically study the ELT in layered superconductors for the simplest case of weak harmonic modulation under conditions when the inequalities $(\omega/c)\sin\theta<k_c<|(\omega/c)\sin\theta\pm g|$ are fulfilled. The transmittance $T$ versus the incidence angle $\theta$ displays a series of resonance peaks with $T\sim1$ emerging by the successive intersection of different dispersion curves for the WGMs with the dispersion curve of the diffracted wave.

\begin{figure}[hbpt]
\includegraphics[width=8.5cm]{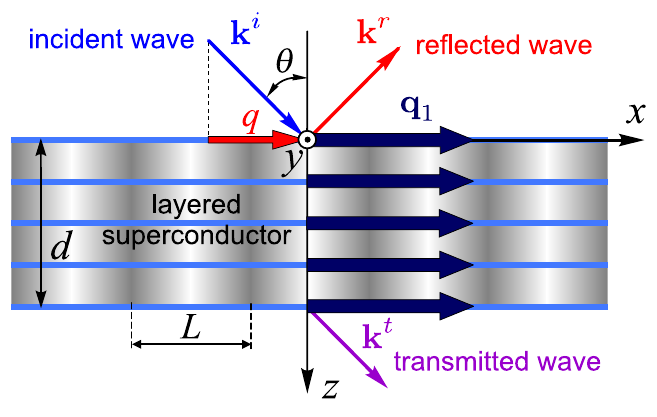}
\caption{(Color online) Schematic geometry of the problem; ${\bf k}^i$, ${\bf k}^r$, and ${\bf k}^t$ are, respectively, the wave vectors of the incident, reflected, and transmitted waves; ${\bf q}_1$ is the longitudinal wave vector of the excited waveguide mode (WGM).}
\label{f1}
\end{figure}

\textit{Electromagnetic field in vacuum and layered superconductor}. ---Consider a slab of layered superconductor of thickness $d$ surrounded by the vacuum (see Fig.~\ref{f1}). The crystallographic $\mathbf{ab}$-plane coincides with the $xy$-plane and the $\mathbf{c}$-axis is directed along the $z$-axis. The plane $z=0$ is the top edge of the slab. Suppose that the maximum $\mathbf{c}$-axis Josephson current density, $J_c$, and, therefore, the Josephson plasma frequency $\omega_J$ are
periodically modulated in the $x$-direction with a spatial period $L$,
\begin{equation}\label{mod}
\omega_J(x)=\omega_J\left[1+f\cos(gx)\right],\quad g=2\pi/L,\quad f\ll1.
\end{equation}

A plane electromagnetic wave of TM (transverse magnetic) polarization,
$\mathbf{E}^{\rm inc}=\left\{E_x^{\rm inc},0,E_z^{\rm inc}\right\}$ and $\mathbf{H}^{\rm inc}=\left\{0,H^{\rm inc},0\right\}$, is incident from the vacuum onto the top side, $z=0$, of the superconductor at an angle $\theta$. The in-plane and out-of-plane components of its wave vector $\mathbf{k}^i$ are
\begin{equation}\label{k}
k_x\equiv q=k\sin\theta,\quad k_z =k\cos\theta,\quad k=\omega/c.
\end{equation}
The in-plane periodic modulation results in generating the diffracted waves with the in-plane wave numbers $q_n=q+ng$. For simplicity, we derive the electromagnetic field distribution taking into account only the first-order diffracted wave, $q_1=q+g$, which provides the ELT under conditions when $q_1$ is close to the wave vector of one of the eigenmodes. Then, the results are generalized to the minus-first diffraction order, $q_{-1}=|q-g|$.

We assume that the diffracted wave in the vacuum is evanescent ($q_1>\omega/c$), and its in-plane wave number $q_1$ is close to the corresponding wave number of one of the eigenmodes. Thus, the electromagnetic field in the vacuum over the superconductor ($z<0$) is presented as a sum of the incident wave (with unit amplitude), specularly diffracted ($n=0$) wave, and the first-order diffracted ($n=1$) evanescent wave. The magnetic $H_{\rm{top}}^V(x,z)$ and the tangential
electric $ E_{x\,\mathrm{top}}^V(x,z)$ fields read
\begin{equation}\label{eq:magn:field:vac:top}
\begin{split}
&H_{\rm{top}}^V(x,z)=\exp(iq x+ikz\cos\theta)\\&+R_{0}\exp(iqx-ikz\cos\theta)+R_{1}\exp(iq_1x+\varkappa_1^Vz),
\end{split}
\end{equation}
\begin{equation}\label{eq:electr:field:vac:top}
\begin{split}
&E_{x\,\mathrm{top}}^V(x,z)=\cos\theta\,[\exp(iqx+ikz\cos\theta)\\
&-R_{0}\,\exp(iqx-ikz\cos\theta)]-\frac{i\varkappa_1^V}{k}\,R_{1}\,\exp(iq_1x+\varkappa_1^Vz),
\end{split}
\end{equation}
where the attenuation coefficient $\varkappa_1^V=\sqrt{q_1^2-k^2}>0$.

The electromagnetic field under the superconductor ($z>d$) is a sum of the transmitted wave and
the first-order diffracted evanescent wave,
\begin{equation}\label{eq:magn:field:vac:bot}
\begin{split}
&H_{\mathrm{bot}}^V(x,z)=T_0\,\exp[iqx+ik(z-d)\cos\theta]\\&+T_{1}\,\exp[iq_1x-\varkappa_1^V(z-d)],
\end{split}
\end{equation}
\begin{equation}\label{eq:electr:field:vac:bot}
\begin{split}
&E_{x\,\mathrm{bot}}^V(x,z)=\cos\theta\,T_0\,\exp[iqx+ik(z-d)\cos\theta]\\&+\frac{i\varkappa_1^V}{k}\,T_{1}\,\exp[iq_1x-\varkappa_1^V(z-d)].
\end{split}
\end{equation}

The electromagnetic field inside the layered superconductor is determined by the distribution of the gauge-invariant phase difference $\varphi(x,z,t)$ of the order parameter between the layers
(see, e.g., Ref.~\cite{Thz-rev}),
\[
\frac{\partial H^s}{\partial x}=-\frac{{\cal H}_0}{\lambda_c\omega_J^2}[\omega_J^2(x)(1-i\Gamma_c)-\omega^2]\varphi,\quad
{\cal H}_0=\frac{\Phi_0}{2\pi D\lambda_c},
\]
\begin{equation}
E^s_x=ik(1+i\Gamma_{ab})\lambda^2_{ab}\frac{\partial H^s}{\partial z},\quad E_z^s=-i{\cal H}_0k\lambda_c\varphi.
\label{sfield}
\end{equation}
Here, $\omega_J(x)$ is given by Eq.~\eqref{mod}, $\omega_J =(8\pi eDJ_c/\hbar\varepsilon)^{1/2}$ is the Josephson plasma frequency unperturbed by the modulation, $J_c$ is the maximum value of the Josephson current density $j_z=J_c\sin\varphi$, $D$ is the spatial period of the layered structure, $\Phi_0=\pi c\hbar/e$ is the magnetic flux quantum. The dimensionless relaxation frequencies $\Gamma_{ab}=4\pi\sigma_{ab}\omega\lambda_{ab}^2/\varepsilon\omega_J^2\lambda_{c}^2$ and $\Gamma_{c}=4\pi\sigma_{c}\omega/\varepsilon\omega_J^2$ are proportional to the averaged quasiparticle conductivities $\sigma_{ab}$ (along the layers) and $\sigma_{c}$ (across the layers). Note that $\Gamma_c$ can actually be neglected due to the smallness of the out-of-plane conductivity $\sigma_c$.

The phase difference $\varphi$ is governed by a set of coupled sine-Gordon equations (see, e.g., review~\cite{Thz-rev} and references therein). For linear JPWs in the continuum limit the coupled sine-Gordon equation can be written as
\begin{equation}\label{eq:SineGord}
\begin{split}
&\left[1-(1+i\Gamma_{ab})\lambda_{ab}^2\frac{\partial^2}{\partial z^2}\right]\left[\omega_{J}^2(x)-\omega^2\right]\varphi\\
&-\lambda_c^2\omega_J^2\frac{\partial^2\varphi}{\partial x^2}=0 .
\end{split}
\end{equation}
Equation \eqref{eq:SineGord} can be solved perturbatively in small modulation amplitude $f\ll1$, see Eq.~\eqref{mod}. Then, Eq.~\eqref{sfield} gives the magnetic field inside the
layered superconductor ($0<z<d$),
\begin{equation}\label{eq:magn:field:sup}
\begin{split}
&H^s(x,z)=\Psi_0(x)\\
&\times\left[C_0^{+}\exp[p_{\,0}(z-d)]+C_0^{-}\exp(-p_{\,0}z)\right]\\
&+\Psi_1(x)\left[C_1^{+}\exp(i\varkappa_1^S z)+C_1^{-}\exp(-i\varkappa_1^S z)\right],
\end{split}
\end{equation}
with
\begin{equation}\label{eq:Psi01}
\begin{split}
\Psi_0(x)=\exp(iqx)-F_{01}\exp(iq_1x)\,,\\\Psi_1(x)=\exp(iq_1x)+F_{01}\exp(iqx)\,;
\end{split}
\end{equation}
\begin{equation}\label{eq:p-kappa-s}
\begin{split}
&p_0=\frac{1}{\lambda_{ab}}\Big(1-\frac{i \Gamma_{ab}}{2}\Big)\sqrt{1-\frac{\lambda_c^2q^2}{\Omega^2-1}}\,\\
&\times\Big(1+\frac{FF_{01}}{2}\frac{q^2}{\Omega^2-1-q^2}\Big)\,,\\
&\varkappa_1^S=\frac{1}{\lambda_{ab}}\Big(1-\frac{i\Gamma_{ab}}{2}\Big)\sqrt{\frac{\lambda_c^2q_1^2}{\Omega^2-1}-1}\,\\
&\times\Big(1+\frac{FF_{01}}{2}\frac{q_1^2}{q_1^2-\Omega^2+1}\Big)\,, \\
& F=\frac{f}{\Omega^2-1},\quad F_{01}=F\frac{qq_1}{q_1^2-q^2},\quad\Omega=\frac{\omega}{\omega_J}.
\end{split}
\end{equation}
The tangential component of the electric field in the layered superconductor reads
\begin{equation}\label{eq:electr:field:sup}
\begin{split}
&E^s_x(x,z)=i\Big(1+\frac{i\Gamma_{ab}}{2}\Big)\\
&\times\Big\{a_0\,\Psi_0(x)\left[C_0^{+}\exp[p_0(z-d)]-C_0^{-}\exp[-p_0z]\right]\,\\
&+a_1\,\Psi_1(x)\left[C_1^{+}\exp[i\varkappa_1^S z]-C_1^{-}\exp[-i\varkappa_1^Sz]\right]\Big\}\,,
\end{split}
\end{equation}
where
\begin{equation}\label{eq:a-01}
a_0=k\lambda_{ab}\sqrt{1-\frac{\lambda_c^2 q^2}{\Omega^2-1}},\quad a_1=k\lambda_{ab}\sqrt{\frac{\lambda_c^2 q_1^2}{\Omega^2-1}-1}
\end{equation}
are the small surface impedances for the basic and the first-order diffracted waves, respectively.

\textit{Transmittance and reflectance}. ---Matching the tangential components of the electric and magnetic fields at the boundaries, $z=0$ and $z=d$, of the superconductor, we obtain eight linear algebraic equations for eight unknown amplitudes, $R_0$, $R_1$, $T_0$, $T_1$, $C_0^+$, $C_0^-$, $C_1^+$, and $C_1^-$. Solving these equations yields the transmittance $|T_0|^2$ and reflectance
$|R_0|^2$ of the superconducting slab,
\begin{eqnarray}
|T_0|^2&=&\frac{4\,F_{01}^4\,a_1^2/\cos^2{\theta}}{\mathcal{D}^2+\mathcal{B}^2}\,,\label{eq:T0:modul2} \\
|R_0|^2&=&\frac{\mathcal{D}^2+\Big(\,\varkappa_1^Sd\,\Gamma_{ab}/2\Big)^2}{\mathcal{D}^2+\mathcal{B}^2}\,,\label{eq:R0:modul2}
\end{eqnarray}
where
\begin{eqnarray}
\mathcal{D}&=&\tan\left(\varkappa_1^S d\,\right)-2\,\frac{k}{\varkappa_1^V}\,a_1\,,\label{eq:D}\\
\mathcal{B}&=&\,\frac{\varkappa_1^S d}{2}\,\Gamma_{ab}+2\,F_{01}^2\,\frac{a_1}{\cos\theta}\,.\label{eq:B}
\end{eqnarray}
Equation $\mathcal{D}=0$ with $f=0$, i.e., with $F=0$ in Eq.~\eqref{eq:p-kappa-s}, defines the spectrum of symmetric and antisymmetric waveguide eigenmodes~\cite{krokhin}. The weak modulation of the plasma frequency \eqref{mod} results in a shift of the dispersion curves and in additional damping, namely, in the leakage of the eigenmode energy due to diffraction. This leakage is specified by the second term in Eq.~\eqref{eq:B}.

\begin{figure}
\includegraphics [width=8 cm]{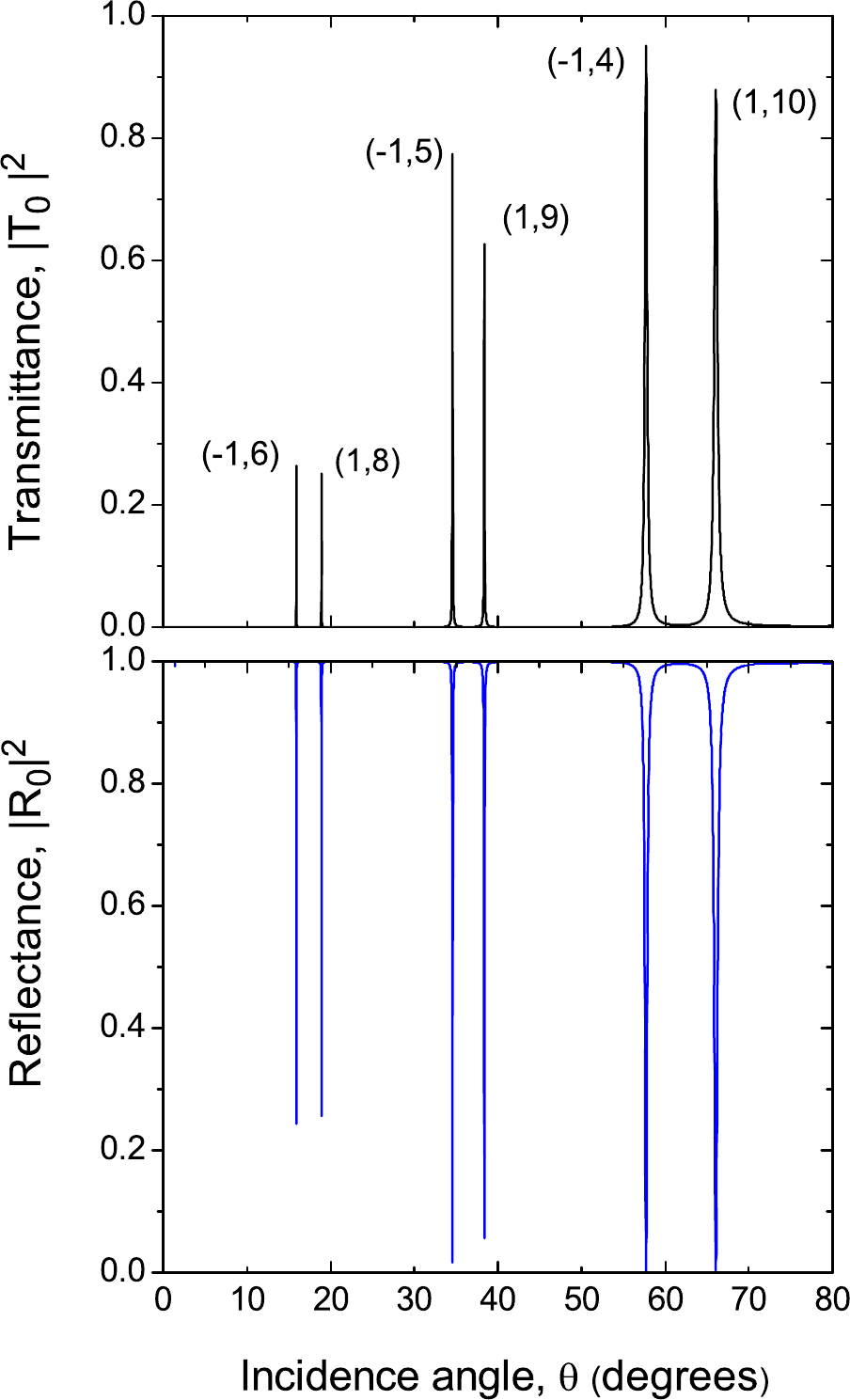}
\caption{\label{f2} (Color online) The transmittance $|T_0|^2$ and reflectance $|R_0|^2$ vs the incidence angle $\theta$ for $d=\lambda_{c}=10\,\lambda_{ab}$, $\varepsilon=16$, $\Gamma_{ab}=10^{-4}$, $f=0.2$, $\Omega=1.1$, $gc/\omega=3$. The first number in the brackets shows the diffraction order responsible for the resonance and the second is the number of the resonantly excited waveguide mode.}
\end{figure}

Equations \eqref{eq:T0:modul2} and \eqref{eq:R0:modul2} describe the resonance enhanced terahertz-wave transmissivity and, respectively,  the suppression of the reflectivity due to the excitation of the waveguide mode by the \emph{first-order} diffracted  wave. Obviously, similar expressions can be derived for the case of the resonance waveguide mode excitation in the \emph{minus-first} diffraction order. To do this, one should substitute the wave number $q_{-1}=|k\sin\theta-g|$ instead of $q_{1}=g+k\sin\theta$ in all formulas. Figure \ref{f2} shows six resonance peaks in the dependence of transmittance $|T_0|^2$ and reflectance $|R_0|^2$ on the incidence angle $\theta$. Three of them are due to the first-order resonance diffraction when the argument $\varkappa_1^S d$ of the tangent in Eq.~\eqref{eq:D} is close to $m\pi$ with $m=8,9,10$; and the others are related to the resonances in the minus-first diffraction order at
$\varkappa_{-1}^S d\approx m\pi$ with $m=4,5,6$.

\begin{figure}
\includegraphics [width=8 cm]{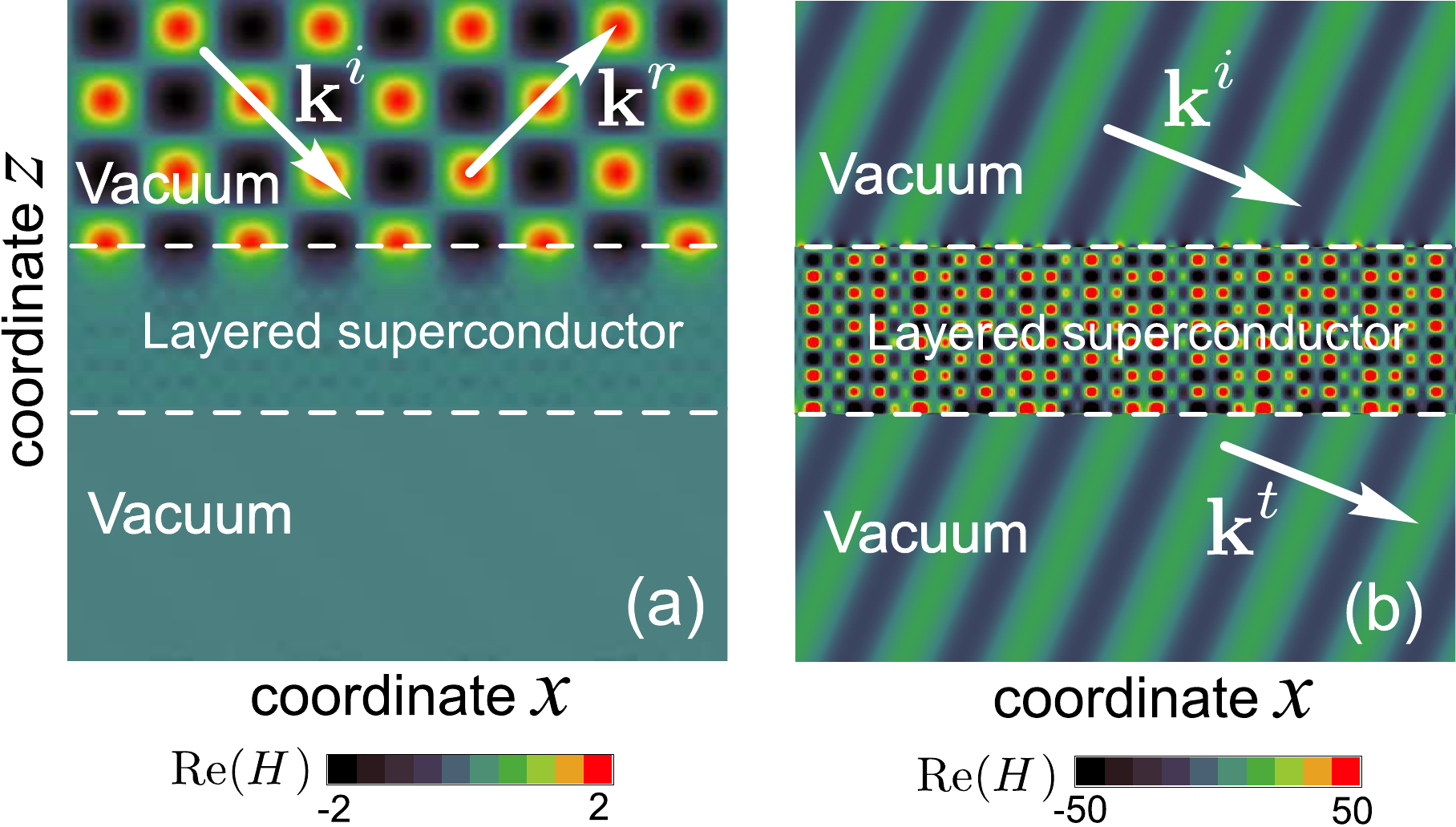}
\caption{\label{f3} (Color online) The magnetic-field distribution: (a) the nonresonance case at $\theta=45^{\circ}$, (b) the resonance diffraction for the order $n=+1$, $\theta=66.1^{\circ}$. The other parameters are the same as in Fig.~\ref{f2}. The superconducting slab is magnified tenfold in the $z$-direction.}
\end{figure}

We also illustrate the ELT effect by the distribution of the magnetic field in Fig.~\ref{f3}. Under the nonresonant conditions, the interference pattern is seen in the vacuum region over the superconductor, and the transmitted wave is absent. In the resonance, the reflected wave is totally suppressed, the interference pattern in the far field disappears, and the transmitted wave is clearly seen under the layered superconductor.

\textit{Conclusions}. ---The enhanced transmissivity of modulated slabs of layered superconductors has been predicted for terahertz radiation. It results from the diffraction of the incident wave and the resonant excitation of the waveguide modes (WGMs). The diffracted wave does not decay deep into the superconducting slab, contrary to metals, where the enhanced light transmission is due to the excitation of evanescent surface waves. A series of resonance peaks, associated with excitation of different WGMs, is observed in the dependence of the transmittance $T$ on the incidence angle $\theta$.

We gratefully acknowledge the support of the SEP-CONACYT (M\'exico) grant No.~166382, Ukrainian State Program on Nanotechnology, and the Program FPNNN of NAS of Ukraine (grant No.~9/12-H).

\end{document}